\documentclass{aa}
\usepackage{graphicx}
\usepackage{txfonts}
\usepackage{orcidlink}
\AtBeginDocument{\nolinenumbers\let\linenumbers\relax}

\def\kps{~km~s$^{-1}$}

\begin{document}

\title{How fast is the fastest-rotating O-type star in the Galaxy?}
\subtitle{LAMOST J040643.69+542347.8}

\author{
M. Curé
\inst{1} \orcidlink{0000-0002-2191-8692}
\thanks{Corresponding author: michel.cure@uv.cl}
\and
R. S. Levenhagen 
\inst{2} \orcidlink{0000-0003-2499-9325}
\and
D. Turis-Gallo
\inst{1} \orcidlink{0009-0000-4835-3479}
\and
G.-W. Li
\inst{3} \orcidlink{0000-0001-7515-6307}
\and
M. Hadjara \orcidlink{0009-0002-6955-8443}
\inst{4,5}
\and
I. Araya
\inst{6} \orcidlink{0000-0002-8717-7858}
\and
C. Arcos
\inst{1} \orcidlink{0000-0002-4825-4910}
}

\institute{
Instituto de F\'{\i}sica y Astronom\'{\i}a, Facultad de Ciencias,
Universidad de Valpara\'{\i}so,
Av. Gran Breta\~na 1111, Valpara\'{\i}so, Chile
\and
Department of Physics, Federal University of Sao Paulo (UNIFESP),
Rua Prof. Artur Riedel 275, 09972-270 Diadema, SP, Brazil
\and
Key Laboratory of Space Astronomy and Technology,
National Astronomical Observatories,
Chinese Academy of Sciences,
Beijing 100101, People's Republic of China
\and
Nanjing Institute of Astronomical Optics \& Technology,
National Astronomical Observatories,
Chinese Academy of Sciences,
Nanjing 210042, China
\and
Chinese High Angular Resolution Southern Astronomical Laboratory
(CHARSAL)/ Department of Astronomy, Universidad de Chile, Camino El Observatorio \#1515, Las Condes, Santiago 7591245, Chile/ Astro-Photonics Laboratory/ Space and Planetary Exploration
Laboratory (SPEL), Dept. of Electrical Engineering, FCFM,
Universidad de Chile, Av. Tupper 2007, Santiago, Chile
\and
Centro Multidisciplinario de F\'{\i}sica,
Vicerrector\'{\i}a de Investigaci\'on,
Universidad Mayor, 8580745 Santiago, Chile
}

\date{Received XXX; accepted YYY}

\abstract 
{Rapid rotation distorts massive stars and produces latitude-dependent effective temperature and gravity. Stellar parameters inferred from a single spherical atmosphere convolved with a rotational profile can therefore be apparent inclination-dependent quantities. LAMOST J040643.69+542347.8 is an extreme Galactic O-type runaway star for which these effects should be appreciable.}
{We re-determined the stellar and rotational parameters of LAMOST J040643.69+542347.8 with an oblate gravity-darkened photospheric model and used selected \ion{He}{ii} transitions treated with tabulated Stark profiles.}
{We fitted continuum-normalised spectra over 4000--4600\,\AA\ with \texttt{ZPEKTR}. The local TLUSTY/SYNSPEC inputs are emergent flux spectra, and we introduced their angular contribution to the visible Roche surface through the limb-darkening treatment implemented in \texttt{ZPEKTR}. The gravity darkening we employed follows the Espinosa-Lara \& Rieutord model, and for the selected \ion{He}{ii} transitions, we used the Sch\"oning--Butler Stark profiles implemented in SYNSPEC. We masked interstellar and wind- or circumstellar-dominated features.} 
{The model reproduces the fitted \ion{He}{ii}~$\lambda4200$, \ion{He}{i}~$\lambda4471$, and \ion{He}{ii}~$\lambda4542$ profiles. The parent non-rotating counterpart (PNRC) parameters are $T_{\rm eff}^{\rm PNRC}=38\,716\pm1041$~K and $\log g^{\rm PNRC}=4.12\pm0.12$, and we obtained $v_{\rm e}\sin i=587\pm73$~\kps. The central grid solution has $\Omega/\Omega_{\rm c}=0.93$, $v_{\rm e}/v_{\rm c}=0.77$, $i=65\degr$, and $R_{\rm e}/R_{\rm p}=1.25$. The corresponding ratio of centrifugal to gravitational acceleration at the equator is $\eta_{\rm e}=0.50$, whereas $\eta_{\rm e}=1$ and $R_{\rm e}/R_{\rm p}=1.5$ at Roche critical rotation. Thus, the angular rotation rate is high, but the central solution is not critical in the equatorial force-balance sense. The projected rotational velocity is higher than but statistically consistent with the value previously obtained from a spherical rotational-convolution analysis.}
{\object{LAMOST J040643.69+542347.8} is an exceptionally rapid O-type rotator with a markedly oblate gravity-darkened photosphere. The ratios $\Omega/\Omega_{\rm c}$, $v_{\rm e}/v_{\rm c}$, and $\eta_{\rm e}$ should be quoted together when discussing its proximity to critical rotation. A controlled spherical fit with the same atmosphere grid would be required to quantitatively isolate the bias introduced by neglecting rotational deformation and gravity darkening.}


\maketitle
\nolinenumbers

\section{Introduction}
\label{sec:introduction}

Rapid rotation affects the structure, evolution, and observed spectra of massive stars. Centrifugal acceleration deforms the stellar surface, while the accompanying variation of effective gravity produces a latitude-dependent radiative flux and effective temperature \citep{vonZeipel1924,EspinosaLaraRieutord2011}. Binary mass transfer or tidal interaction can also spin up a massive star, and a subsequent disruption may leave a rapidly rotating runaway object \citep{Packet1981,Sana2012,deMink2013}. These processes are relevant to \object{LAMOST~J040643.69+542347.8}, which is both an extreme rotator and a runaway star \citep{Li2020}.

The spectroscopic consequences of rotational deformation have been recognised for many years. \citet{Stoeckley1968} showed that departures from spherical symmetry, including gravity darkening, can introduce systematic errors in projected rotational velocities inferred from line widths. \citet{Fremat2005} later developed a quantitative framework linking the parameters obtained from plane-parallel fits, which they termed `apparent parameters', to those of the parent non-rotating counterpart (PNRC) of a rapidly rotating star. The reduced contribution of the cool, faint equatorial zones can make rotationally broadened profiles from a uniformly bright spherical surface narrower than expected \citep{Stoeckley1968,Townsend2004,Fremat2005}.

\citet{Li2020} classified \object{LAMOST~J040643.69+542347.8} as O6.5~Vnnn(f)p. From Potsdam Wolf-Rayet (PoWR) models spectra convolved with a classical rotational profile, they found $T_{\rm eff}=35$~kK and $\log g=3.6$ for their best Large Magellanic Cloud (LMC) metallicity model, with $v_{\rm e}\sin i=543\pm29~\mathrm{km\,s^{-1}}$. Their best solar metallicity model gave $T_{\rm eff}=34$~kK, $\log g=3.6$, and $v_{\rm e}\sin i=538\pm29~\mathrm{km\,s^{-1}}$. These quantities are apparent parameters in the sense of \citet{Fremat2005}, as they describe a spherical one-dimensional fit over a finite spectral interval and need not coincide with either local photospheric conditions or PNRC parameters. \citet{Li2020} also noted that gravity darkening could lead to an underestimate of $v_{\rm e}\sin i$.

In this work we analyse the same low-resolution LAMOST spectrum with the \texttt{ZPEKTR} framework \citep{Levenhagen2014,Levenhagen2024} using a Roche-distorted surface and the gravity-darkening prescription of \citet{EspinosaLaraRieutord2011}. The local atmosphere grid is based on TLUSTY/SYNSPEC \citep{HubenyLanz1995,HubenyLanz2011,HubenyLanz2017}, with selected \ion{He}{ii} transitions evaluated using the Sch\"oning--Butler Stark profiles \citep{SchoeningButler1989}. Our aim was to determine a physically consistent rotating-star solution and clearly distinguish local, apparent, and PNRC quantities. We do not use the comparison with \citet{Li2020} as a controlled test of method performance because an otherwise identical spherical fit with the same atmosphere grid and fitting procedure is not available.

\section{Methodology}
\label{sec:methodology}
\subsection{ZPEKTR code framework}
\label{sec:zpektr_framework}
The \texttt{ZPEKTR} code framework was developed to account for limb darkening, gravity darkening, and the geometrical deformation of rapidly rotating stars \citep{Levenhagen2014,Levenhagen2024}. To compute the disk-integrated photospheric spectrum, ZPEKTR divides the visible Roche surface into finite elements, each characterised by its local radius, effective gravity, effective temperature, projected area, line-of-sight velocity, and viewing angle. The atmosphere files used in the present calculation provide local emergent flux spectra. Direction-dependent contributions to the observer are then constructed within \texttt{ZPEKTR} through its limb-darkening treatment before the Doppler-shifted surface elements are integrated. This distinction is important, as the local angle-dependent quantities entering the surface integral are not direct angle-resolved TLUSTY/SYNSPEC intensities from the grid used here.

\subsection{Rotational geometry and surface discretisation}
\label{sec:geometry}

Within the Roche approximation, the star rotates uniformly, and we treat its mass as centrally concentrated, so the gravitational field is that of a point mass. The stellar surface then coincides with the equipotential of the effective potential in the co-rotating frame \citep{EspinosaLaraRieutord2011}:
\begin{equation}
\Phi(r,\theta) =
-\frac{GM}{r}
-\frac{1}{2}\Omega^{2}r^{2}\sin^{2}\theta\, ,
\label{eq:roche_potential}
\end{equation}
where $G$ is the gravitational constant, $M$ is the stellar mass, $r$ is the radial coordinate from the stellar centre, $\Omega$ is the angular velocity, and $\theta$ is the co-latitude measured from the rotation axis. The degree of rotation is reported through $\Omega/\Omega_{\mathrm{c}}$, where $\Omega_{\mathrm{c}}$ denotes the critical angular velocity at which the effective gravity vanishes at the equator. In the grid used here, $R_{\rm p}$ and $R_{\rm e}$ were sampled directly. Their ratio $x=R_{\rm e}/R_{\rm p}$ fixes the rotational deformation, and within the Roche model, it determines the corresponding angular velocity ratio through \citep[derived from][]{2002A&A...393..345D,2014A&A...569A..45H}
\begin{equation}
\frac{\Omega}{\Omega_{\rm c}}=\left[\frac{27}{4}\,\frac{x-1}{x^{3}}\right]^{1/2}.
\label{eq:angular-velocity ratio}
\end{equation}
The equipotential condition $\Phi(R(\theta),\theta)=\Phi(R_{\rm p},0)$ then determines the latitude-dependent surface radius $R(\theta)$. Thus, the pole-to-equator radius ratio $R_{\rm p}/R_{\rm e}$ and $\Omega/\Omega_{\rm c}$ are linked by the Roche geometry rather than being independent grid coordinates.

A useful measure of the equatorial force balance is
\begin{equation}
\eta_{\rm e}\equiv
\frac{\Omega^{2}R_{\rm e}^{3}}{GM}
=2\left(\frac{R_{\rm e}}{R_{\rm p}}-1\right),
\label{eq:eta}
\end{equation}
where the second equality follows directly from the Roche equipotential condition between pole and equator. Thus, a critical Roche rotator has $\eta_{\rm e}=1$ and $R_{\rm e}/R_{\rm p}=1.5$. We quote this parameter alongside $\Omega/\Omega_{\rm c}$ and $v_{\rm e}/v_{\rm c}$ below since these three ratios do not approach unity at the same rate.

For numerical integration, we discretised the Roche surface into a sufficiently fine set of finite surface elements to ensure accuracy. The adopted sampling resolves the latitude-dependent variations in the radius, effective gravity, effective temperature, projected area, and line-of-sight velocity of the star. The surface resolution and integration scheme followed the convergence tests in \citet{Levenhagen2024} and remained fixed throughout the analysis.

For each surface element, the \texttt{ZPEKTR} code determines the co-latitude, azimuth, outward unit normal, surface area, projected area, and visibility to the observer. The projection factor is defined as $\mu=\hat{\mathbf{n}}\cdot\hat{\mathbf{k}}$, where $\hat{\mathbf{n}}$ is the outward unit vector normal to the local surface and $\hat{\mathbf{k}}$ is the unit vector directed towards the observer. An element contributes to the emergent spectrum only when $\mu>0$; in that case, its projected area is $\mu\,\Delta A$, where $\Delta A$ is the element's surface area.

\subsection{Gravity darkening and local photospheric parameters}
\label{sec:gravity_darkening}

It is important to distinguish between gravitational acceleration and effective gravity. At co-latitude $\theta$, the magnitude of the former is
\begin{equation}
 g_{\rm grav}(\theta)=\frac{GM}{R(\theta)^{2}}\, ,
 \label{eq:gravitational_gravity}
\end{equation} whereas the effective-gravity vector follows from the Roche potential
\begin{equation}
 \boldsymbol{g}_{\rm eff}(\theta)=-\boldsymbol{\nabla}\Phi.
 \label{eq:effective_gravity}
\end{equation}

The latter includes both gravitational and centrifugal accelerations, and both their magnitude and direction vary with co-latitude. At the equator, the effective-gravity magnitude reduces to
\begin{equation}
 g_{{\rm eff},{\rm e}}
 =\frac{GM}{R_{\rm e}^{2}}-\Omega^{2}R_{\rm e}.
 \label{eq:equatorial_effective_gravity}
\end{equation}
Effective gravity is the quantity used in the gravity-darkening prescription and in selecting the local TLUSTY/SYNSPEC spectrum. We report the values of $GM/R^{2}$ separately below as the gravitational acceleration.

The \texttt{ZPEKTR} code can convert the resulting surface distribution of effective gravity into a corresponding effective temperature distribution using either of two gravity-darkening prescriptions. The first is the classical von Zeipel model \citep{vonZeipel1924}:
\begin{equation}
T_{\mathrm{eff}}(\theta)=\Gamma\,
\left[g_{\mathrm{eff}}(\theta)\right]^{\beta},
\label{eq:von_zeipel}
\end{equation} where $\Gamma$ is a normalisation constant that sets the overall temperature scale and $\beta$ is the gravity-darkening exponent. For a star with a radiative envelope, the classical value is $\beta=1/4$, corresponding to a local radiative flux ($\mathbf{F}$) proportional to the effective gravity ($\mathbf{F} \propto \mathbf{g_{\mathrm{eff}}}$).

The second prescription is that of \citet{EspinosaLaraRieutord2011}. Rather than imposing a fixed power-law relation between $T_{\mathrm{eff}}$ and $g_{\mathrm{eff}}$, this formulation derives the surface flux distribution by assuming that the radiative flux vector is locally anti-parallel to the effective gravity vector and satisfies $\nabla\cdot\mathbf{F}=0$ outside the energy-generating core. By explicitly accounting for the rotationally distorted geometry, it generally predicts a weaker pole-to-equator temperature contrast than the classical von Zeipel model at high rotation rates \citep{Levenhagen2024}.

The Espinosa-Lara \& Rieutord model should not be interpreted as Eq.~(\ref{eq:von_zeipel}) with a constant $\Gamma$ and $\beta$. If its surface temperature distribution is represented locally by a power law, the corresponding effective exponent and normalisation depend on latitude. A first-order expansion can be expressed through an effective coefficient \citep{Rieutord2016}, but that coefficient is only a convenient approximation to the full gravity-darkening law. Interferometric determinations also show departures from the canonical $\beta=0.25$; for example, \citet{Domiciano2014} obtained $\beta=0.166^{+0.012}_{-0.010}$ for Achernar.

All results reported in this work were obtained using the gravity-darkening prescription of \citet{EspinosaLaraRieutord2011}, which provides, for each surface element, the local $T_{\rm eff}(\theta)$ associated with $g_{\rm eff}(\theta)$. The parameters adopted to construct the model grid are listed in Table~\ref{tab:grid}. The polar and equatorial temperatures and gravities reported below are therefore local properties of the deformed rotating star and must be distinguished from $T_{\rm eff}^{\rm PNRC}$ and $\log g^{\rm PNRC}$. We did not perform an otherwise identical refit with the von Zeipel prescription, and consequently, the sensitivity to this choice is a source of systematic uncertainty that was not included in the quoted errors.

\begin{table}[!ht]
\centering
\caption{Parameter ranges of the \texttt{ZPEKTR} model grid.}
\label{tab:grid}
\begin{tabular}{lrrr}
\hline \hline
Parameter & Minimum & Maximum & Step \\
\hline
$T_{\rm pole}$ [kK]  & 33.0 & 50.00 & 1.0 \\
$M\ [M_\odot]$       & 10.0 & 55.00 & 5.0 \\
$R_p\ [R_\odot]$     & 10.0 & 26.00 & 2.0 \\
$R_e\ [R_\odot]$     & 12.5 & 38.12 & non-uniform$^{a}$ \\
$i$ [deg]            &  5.0 & 90.00 & 5.0 \\
\end{tabular}
\tablefoot{Total models: 223\,560\\
$^{a}$ Non-uniform sampling in $R_e$.}
\end{table}

\subsection{Local atmosphere spectra and the Sch\"oning--Butler profiles}
\label{sec:local_spectra}

Each visible surface element was assigned a local synthetic spectrum by interpolation in the plane-parallel TLUSTY/SYNSPEC grid \citep{HubenyLanz1995,HubenyLanz2011,HubenyLanz2017}. The files used for the present grid are emergent flux spectra, as described for the \texttt{ZPEKTR} input grids by \citet{Levenhagen2024}, rather than angle-resolved specific intensities. For the surface integration, we represent the local angular dependence with a limb-darkening function, ${\cal L}_{\lambda}(\mu)$. By writing $I_{\lambda}(\mu)=I_{\lambda}(1){\cal L}_{\lambda}(\mu)$, its normalisation is fixed by
\begin{equation}
F_{\lambda}=2\pi I_{\lambda}(1)\int_{0}^{1}{\cal L}_{\lambda}(\mu)\,\mu\,{\rm d}\mu .
\label{eq:flux_to_intensity}
\end{equation}
Accordingly, the local $I_{\lambda}$ entering the disk integral below is reconstructed from the atmospheric flux and the adopted limb-darkening relation; it is not a direct SYNSPEC specific-intensity output. \texttt{ZPEKTR} implements linear, quadratic, square-root, and logarithmic laws \citep{Levenhagen2014,Levenhagen2024}.

For the selected \ion{He}{ii} transitions, SYNSPEC uses tabulated Stark-broadened profiles calculated by \citet{SchoeningButler1989}. Their implementation in the spectrum synthesis code is described by \citet{HubenyLanz2000,HubenyLanz2017}. The profiles are based on the unified theory of \citet{VidalCooperSmith1970}, modified to account for the Coulomb deflection of electrons by the charged \ion{He}{ii} radiator. The electron contribution is treated continuously from the impact regime in the line core to the quasi-static regime in the wings, whereas the perturbing ions are assumed to be quasi-static; therefore, the ion-dynamical effects are neglected.  SYNSPEC interpolates the thermally Doppler-convolved profiles at each atmospheric depth as a function of local temperature and electron density. In particular, \ion{He}{ii}~$\lambda4542$ ($n=4\rightarrow9$) is evaluated from the tabulated depth-dependent profile rather than being represented by an additional Lorentzian damping term or by post-processing convolution. Throughout the analysis presented here, the selected \ion{He}{ii} transitions were treated consistently with the dedicated Sch\"oning--Butler Stark-profile tabulations implemented in SYNSPEC.

\subsection{Disk integration and spectral fitting}
\label{sec:integration_fitting}

The monochromatic flux seen at the inclination angle $i$ was obtained by summing the Doppler-shifted specific intensities of all visible elements:
\begin{equation}
 F_{\lambda}(i) \propto
 \sum_{j,\,\mu_j>0}
 I_{\lambda,j}\!\left(T_{{\rm eff},j},g_{{\rm eff},j},\mu_j\right)
 \mu_j\,\Delta A_j,
 \label{eq:disk_integration}
\end{equation}
where $\Delta A_j$ is the area of element $j$. We shifted the local wavelength by the element's projected rotational velocity before evaluating the surface sum. We applied the same integration to the local continuum, which allowed us to normalise the final model spectrum consistently.

We analysed the same low-resolution LAMOST optical spectrum used by \citet{Li2020} over the wavelength interval $4000$--$4600$\,\AA. The nominal resolving power of the LAMOST low-resolution spectra is $R\simeq1800$, and we convolved the model to the corresponding instrumental resolution before comparing it with the continuum-normalised observations. We excluded diffuse interstellar bands used from the photospheric fit as well as regions whose profiles were dominated by circumstellar or wind emission. \ion{He}{i}~$\lambda4471\, \mbox{\AA}\,$ was retained as a photospheric diagnostic because this line is only weakly affected by wind and circumstellar emission in rapidly rotating Be-star applications \citep{Levenhagen2014}; its observed asymmetry is mainly associated with the nearby forbidden \ion{He}{i}~$\lambda4470\, \mbox{\AA}\,$ component rather than with wind broadening. The principal fitted photospheric diagnostics include the optical \ion{He}{ii} transitions, with particular emphasis on \ion{He}{ii}~$\lambda4542\, \mbox{\AA}$. 

The \texttt{ZPEKTR} grid was constructed by sampling $T_{\rm pole}$, $M$, $R_{\rm p}$, $R_{\rm e}$, and $i$. We used the PNRC notation as introduced by \citet{Fremat2005}. The PNRC parameters describe the non-deformed counterpart with the same mass, whereas `apparent' parameters are those obtained by fitting the observed rotating star with classical plane-parallel models. For each \texttt{ZPEKTR} model, $T_{\rm eff}^{\rm PNRC}$, $R_{\rm PNRC}$, $\log g^{\rm PNRC}$, and $\Omega/\Omega_{\rm c}$ were derived from the adopted geometry and physical relations. They are therefore not independent grid coordinates. Before evaluating the goodness of fit, we convolved each \texttt{ZPEKTR} model spectrum with the LAMOST instrumental profile described by \citet{Li2020}. The goodness of fit was evaluated through a relative $\chi^{2}$-like merit function over the selected spectral pixels,
\begin{equation}
\chi^{2}_{\rm rel} = \frac{1}{N_{\mathrm{lin}}} \sum_{k=1}^{N_{\mathrm{lin}}}
\frac{1}{n_{k}} \sum_{i=1}^{n_{k}}
\frac{\left[ F^{\mathrm{obs}}_{k}(\lambda_i)
- F^{\mathrm{mod}}_{k}(\lambda_i) \right]^{2}}{F^{\mathrm{mod}}_{k}(\lambda_i)} \, ,
\label{eq:chi2}
\end{equation}where $F^{\mathrm{obs}}_{k}(\lambda_i)$ is the observed continuum-normalised flux and $F^{\mathrm{mod}}_{k}(\lambda_i)$ denotes the continuum-normalised \texttt{ZPEKTR} model flux after convolution with the LAMOST instrumental profile, $N_{\mathrm{lin}}$ is the number of fitted lines, and $n_{k}$ is the number of points in the spectral window of line $k$. In the present fit, $N_{\mathrm{lin}}=3$, corresponding to \ion{He}{ii}~$\lambda4200$, \ion{He}{i}~$\lambda4471$, and \ion{He}{ii}~$\lambda4542$. The $1/n_{k}$ factor prevents wider windows from dominating the fit such that all lines contribute equally, regardless of their window width. Because Eq.~(\ref{eq:chi2}) is a relative merit function whose denominator is not an observational variance, we did not interpret its numerical minimum as a conventional reduced $\chi^{2}$ or assign it an absolute goodness-of-fit probability.

We ranked the models by their $\chi^{2}_{\rm rel}$ values. The reported central solution
is the best-fitting grid model, while the quoted uncertainties are the
standard deviations of the parameter distributions within the first quartile
of the selected models \citep[see details in][]{Turis-Gallo2025}. These dispersions describe the adopted model-selection
procedure and do not include systematic uncertainties associated with the
gravity-darkening prescription, continuum placement, or grid boundaries.
For parameters whose solutions accumulate at a physical or grid boundary,
the selected-model distribution is truncated, and a symmetric uncertainty
must not be interpreted as a two-sided confidence interval.

Because spectroscopy measures only the disk-integrated stellar flux, it is insensitive to the position angle of the stellar rotation axis ($PA_{\rm rot}$) in the plane of the sky. Interferometric phase measurements provide access to this quantity, enabling studies of the orientation of stellar angular momentum with respect to proper motions, circumstellar structures, and the parent star-forming environment \citep[][]{2018MNRAS.480.1263H,2022MNRAS.511.4724H}.

\section{Results}
\label{sec:results}

\subsection{Spectral fit with the Sch\"oning--Butler \ion{He}{ii} profiles}
\label{sec:stark_results}

\begin{figure*}[t]
    \centering
    \includegraphics[width=0.96\linewidth]{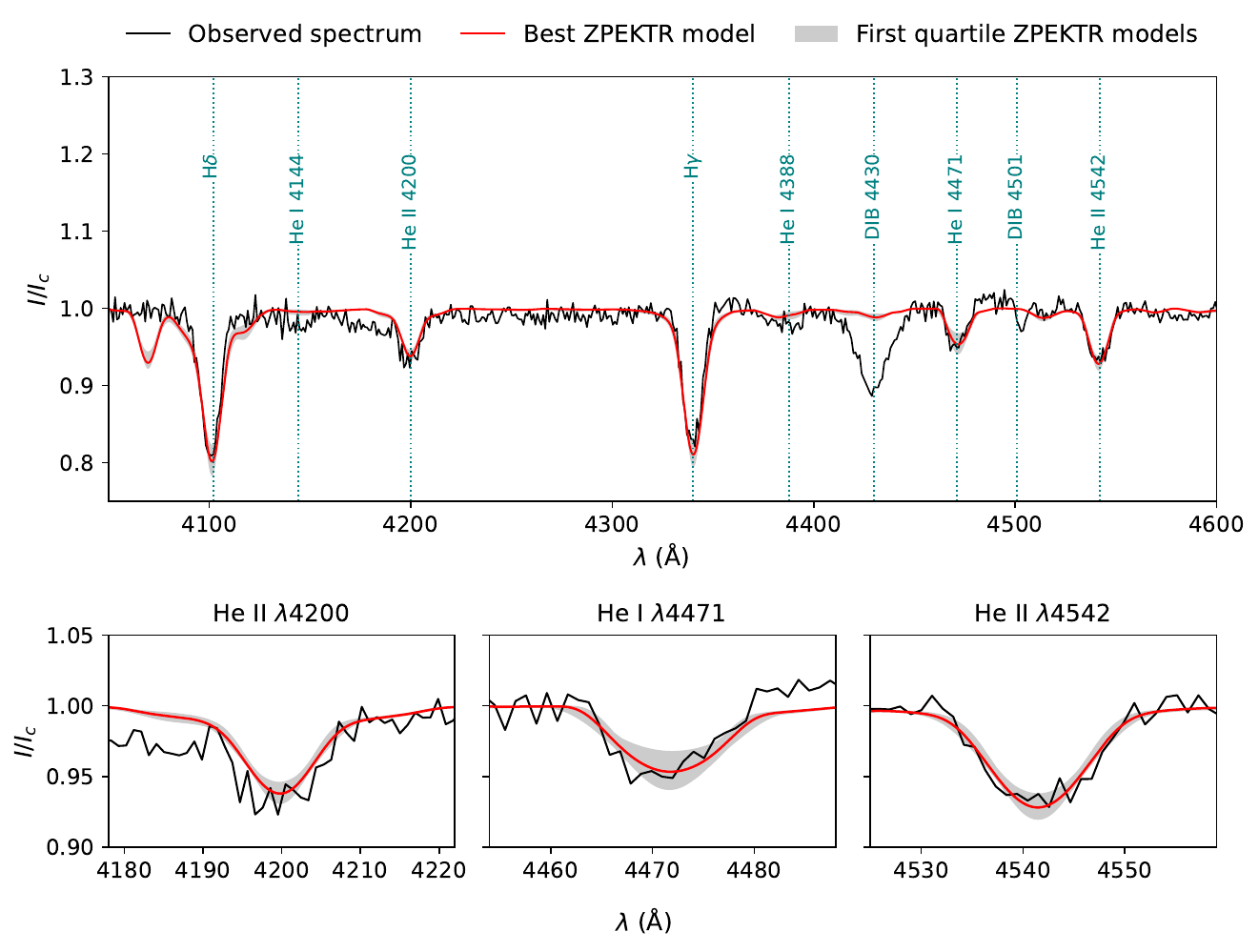}
    \caption{Comparison between the observed spectrum of \object{LAMOST J040643.69+542347.8} (black) and the best-fitting \texttt{ZPEKTR} model (red). Top panel: Full displayed spectral range 4050--4600~\AA. The shaded grey area indicates the range spanned by the first quartile of the $\chi^{2}_{\rm rel}$ distribution, and the dotted teal lines mark the main spectral features. Bottom panels: Close-up view of the three lines used in the fit, He\,{\sc ii}~$\lambda$4200, He\,{\sc i}~$\lambda$4471, and He\,{\sc ii}~$\lambda$4542.}
    \label{fig:zpektr_fit}
\end{figure*}

The selected \ion{He}{ii} transitions were synthesised using the tabulated Sch\"oning--Butler Stark profiles. Figure~\ref{fig:zpektr_fit} shows a comparison of the observed LAMOST spectrum with the best \texttt{ZPEKTR} model obtained using this grid. The model reproduces the principal fitted helium diagnostics over $4000$--$4600$~\AA, while the prominent diffuse interstellar absorption near $4430$~\AA\ and the other masked interstellar features are intentionally not fitted. The lower panels highlight the agreement between the observed and modelled profiles of \ion{He}{ii}~$\lambda4200$, \ion{He}{i}~$\lambda4471$, and \ion{He}{ii}~$\lambda4542$. The grey band represents the spread of the first quartile of the models selected in this analysis. Table~\ref{tab:bestfit_parameters} provides a summary of the stellar and rotational parameters obtained.

\begin{table}
 \caption{Stellar and rotational parameters.}
 \label{tab:bestfit_parameters}
 \centering
 \begin{tabular}{lc}
  \hline\hline
  Parameter & Sch\"oning--Butler solution \\
  \hline
   $T_{\rm eff}^{\rm PNRC}$ (K) & $38\,716 \pm 1041$ \\
  $\log g^{\rm PNRC}$ (cgs) & $4.12 \pm 0.12$ \\
  $v_{\rm e}\sin i$ (km\,s$^{-1}$) & $587 \pm 73$ \\
  $M/M_{\odot}$ & $55.0 \pm 8.9$ \\
  $R^{\rm PNRC}/R_{\odot}$ & $10.8 \pm 1.1$ \\
  $\log(L^{\rm PNRC}/L_{\odot})$ & $5.377 \pm 0.076$ \\
  $\Omega/\Omega_{\rm c}$ & $0.93 \pm 0.05 $ \\
  $v_{\rm e}/v_{\rm c}$ & $0.77 \pm 0.13$ \\
  $\eta_{\rm e}$ & $0.50$ \\
  $R_{\rm e}/R_{\rm p}$ & $1.25$ \\
  $i$ (deg) & $65 \pm 12$  \\
  $T_{\rm p}$ (K) & $41\,000 \pm 1113$ \\
  $T_{\rm e}$ (K) & $34\,125 \pm 4040$ \\
  $\log g_{{\rm grav},{\rm p}}$ (cgs) & $4.18 \pm 0.12$ \\
  $\log g_{{\rm grav},{\rm e}}$ (cgs) & $3.98 \pm 0.18$ \\
  $R_{\rm p}/R_{\odot}$ & $10.0 \pm 1.0$ \\
  $R_{\rm e}/R_{\odot}$ & $12.5 \pm 2.2$ \\
  \hline
 \end{tabular}
 \tablefoot{Parameters were obtained from the \texttt{ZPEKTR} fit using
the Sch\"oning--Butler \ion{He}{ii} Stark profiles. PNRC quantities refer
to the parent non-rotating counterpart \citep{Fremat2005}; polar and
equatorial quantities refer to the deformed model. The tabulated
$g_{\rm grav}=GM/R^{2}$ values exclude centrifugal acceleration. The force
ratio $\eta_{\rm e}$ and the flattening $R_{\rm e}/R_{\rm p}$ are quoted
for the central grid solution; a critical Roche rotator has
$\eta_{\rm e}=1$ and $R_{\rm e}/R_{\rm p}=1.5$.
}
\end{table}

\subsection{Stellar and rotational parameters}
\label{sec:stellar_parameters}

Table~\ref{tab:bestfit_parameters} contains the fitted and derived quantities. We emphasise only the aspects needed for the discussion below. The central model has $\Omega/\Omega_{\rm c}=0.93$, $v_{\rm e}/v_{\rm c}=0.77$, and $R_{\rm e}/R_{\rm p}=1.25$, and hence $\eta_{\rm e}=0.50$ through Eq.~(\ref{eq:eta}). Thus, although the angular-frequency ratio is high, the equatorial centrifugal acceleration is only half the gravitational acceleration in the central model. The central values of $M$, $R_{\rm p}$, and $R_{\rm e}$ lie at adopted grid boundaries, so their quoted dispersions are internal boundary-truncated measures rather than two-sided confidence intervals.

For reference, Table~\ref{tab:apparent_comparison} lists the apparent parameters \citet{Li2020} obtained from spherical PoWR spectra. We include them to explicitly distinguish between apparent and PNRC quantities, not as a controlled comparison of fitting methods. As \citet{Fremat2005} stressed, apparent parameters inferred from a spherical analysis of a rapid rotator depend on the spectral diagnostics and wavelength interval used in the fit.

\begin{table}[!ht]
\centering
\caption{Apparent parameters from a previous spherical analysis.}
\label{tab:apparent_comparison}
\begin{tabular}{lcc}
\hline\hline
Parameter & LMC grid & Solar grid \\
\hline
$T_{\rm eff}$ (kK) & 35 & 34 \\
$\log g$ (cgs) & 3.6 & 3.6 \\
$v_{\rm e}\sin i$ (km\,s$^{-1}$) & $543\pm29$ & $538\pm29$ \\
\hline
\end{tabular}
\tablefoot{The values are from \citet{Li2020}.\\
The best LMC- and solar-metallicity PoWR models were fitted to \ion{He}{ii}~$\lambda4542$ with a classical rotational convolution.}
\end{table}

For comparison with older power-law descriptions only, the first term of the expansion discussed by \citet{Rieutord2016} may be written as
\begin{equation}
\beta_{\rm eff}\simeq \frac{1}{4}
-\frac{1}{3}\left(1-\frac{R_{\rm p}}{R_{\rm e}}\right).
\label{eq:beta_effective}
\end{equation}
For the central ratio $R_{\rm p}/R_{\rm e}=0.80$, this gives $\beta_{\rm eff}\simeq0.18$. We stress that Eq.~(\ref{eq:beta_effective}) is not the gravity-darkening law used in the synthesis and that $\beta_{\rm eff}$ is not a fitted constant of the Espinosa-Lara \& Rieutord model.

\section{Discussion}
\label{sec:discussion}

\subsection{Spectral fit and local line physics}
\label{sec:discussion_stark}

The Sch\"oning--Butler profiles were used consistently for the selected \ion{He}{ii} transitions. This is relevant for \ion{He}{ii}~$\lambda4542$, whose wings respond to the local electron density and gravity. In a gravity-darkened model, the disk-integrated line combines surface elements covering a broad range of $T_{\rm eff}$ and $g_{\rm eff}$, so the local profile should respond to the atmosphere assigned to each element. In the present work, we do not compare alternative Stark-broadening prescriptions, and therefore the remaining spectral discrepancies cannot be attributed uniquely to Stark broadening.

\subsection{Rotation rate and degree of criticality}
\label{sec:discussion_rotation}

The three dimensionless rotation measures used here describe different aspects of the same Roche model. The central solution has $\Omega/\Omega_{\rm c}=0.93$, but $v_{\rm e}/v_{\rm c}=0.77$ and $\eta_{\rm e}=0.50$. Its equatorial-to-polar radius ratio is $R_{\rm e}/R_{\rm p}=1.25$, compared with 1.5 at the Roche critical rotation. We therefore describe \object{LAMOST J040643.69+542347.8} as an exceptionally rapid rotator with a high angular-frequency ratio, rather than using ``near-critical'' without qualification.

Gravity darkening remains important even though $\eta_{\rm e}<1$. The equatorial zones carry the largest line-of-sight rotational velocities but contribute less flux than hotter intermediate and polar latitudes. \citet{Stoeckley1968} identified this effect in line-profile calculations, and later work quantified it \citep{Townsend2004,Fremat2005}. The difference between apparent and true $v_{\rm e}\sin i$ is not universal. It depends on the line, inclination, temperature, gravity, and stellar mass. In particular, \citet{Fremat2005} discussed the dependence of rotational deformation on the ratio $\tau=K/|W|$ of rotational kinetic to gravitational potential energy and showed the corresponding mass dependence of the $v\sin i$ corrections (their Fig.~5). This is relevant here because the present object is much more massive than B stars, which are often used to illustrate the largest saturation effects.

\subsection{Apparent and PNRC parameters}
\label{sec:discussion_li2020}

The value $v_{\rm e}\sin i=587\pm73~\mathrm{km\,s^{-1}}$ is statistically consistent with the \citet{Li2020} measurements in Table~\ref{tab:apparent_comparison}. Likewise, $T_{\rm eff}^{\rm PNRC}=38\,716\pm1041$~K and $\log g^{\rm PNRC}=4.12\pm0.12$ should not be compared as if they were the same observables as the apparent $T_{\rm eff}$ and $\log g$ from a spherical fit. The rotating model instead contains a surface distribution, from $T_{\rm e}\simeq34.1$~kK to $T_{\rm p}\simeq41.0$~kK, together with a corresponding range of effective gravities.

This distinction also limits what we can conclude from the present comparison. Over a restricted wavelength interval, a spherical rotational-convolution model can provide an excellent formal fit even when rotational deformation and gravity darkening are physically important; the fitted quantities are then apparent parameters tied to the diagnostics being analysed \citep{Fremat2005}. A strict test of the rotational effect in the present case would therefore require fitting the same wavelength interval with the same TLUSTY/SYNSPEC grid, masks, instrumental convolution, merit function, and optimisation procedure, changing only the surface treatment from the oblate gravity-darkened calculation to a spherical rotational convolution. Such a same-grid control fit is not contained in the calculations available here. We therefore do not claim that the numerical difference relative to \citet{Li2020} demonstrates the superiority of one fitting method. The present result instead provides a physically explicit rotating-star interpretation.

The \ion{He}{ii}~$\lambda4686$ and H$\alpha$ profiles mentioned by \citet{Li2020} lie outside the 4000--4600~\AA\ interval fitted here. They are shown in their Fig.~2c, and \ion{He}{ii}~$\lambda4686$ also varies among the available low-resolution spectra. We therefore did not add these wind-sensitive lines to the photospheric fit. Their broad emission and central reversals may reflect an anisotropic wind or circumstellar material, and they require treatment beyond the present surface-integration model.

\subsection{Model limitations}
\label{sec:discussion_limitations}

The LAMOST spectrum has $R\simeq1800$, which limits the information available in the line cores and makes continuum placement and interstellar contamination important. The model grid also reaches boundaries in $M$, $R_{\rm p}$, and $R_{\rm e}$. The quoted dispersions are consequently internal to the adopted grid and do not include systematic uncertainties from the gravity-darkening prescription, limb-darkening prescription, continuum normalisation, or alternative atmosphere grids.

Two controlled calculations would be especially useful: a refit with a different gravity-darkening prescription and the same-grid spherical comparison described above. High-resolution multi-epoch spectroscopy would further help separate photospheric, wind, and circumstellar contributions and test the stability of the line asymmetries.

\section{Conclusions}
\label{sec:conclusions}

We have analysed \object{LAMOST J040643.69+542347.8} with an oblate gravity-darkened \texttt{ZPEKTR} model and TLUSTY/SYNSPEC local spectra using the Sch\"oning--Butler Stark profiles for the selected \ion{He}{ii} transitions. The fitted profiles were reproduced with $v_{\rm e}\sin i=587\pm73~\mathrm{km\,s^{-1}}$ and PNRC parameters $T_{\rm eff}^{\rm PNRC}=38\,716\pm1041$~K and $\log g^{\rm PNRC}=4.12\pm0.12$.

The central model has $\Omega/\Omega_{\rm c}=0.93$, $v_{\rm e}/v_{\rm c}=0.77$, $R_{\rm e}/R_{\rm p}=1.25$, and $\eta_{\rm e}=0.50$. These values show that the star is strongly deformed and rotates at a high fraction of the critical angular frequency but is not at a critical equatorial force balance. This distinction matters when characterising its degree of criticality.

The results of the comparison we performed with the spherical analysis of \citet{Li2020} are consistent with the expected influence of rotational deformation and gravity darkening. Still, it was not a controlled one-to-one test because the atmosphere grids and fitting procedures differ. A spherical fit using the same atmosphere grid and fitting procedure would be required to quantify any bias in the inferred projected rotational velocity caused by neglecting rotational deformation and gravity darkening. The wind-sensitive \ion{He}{ii}~$\lambda4686$ and H$\alpha$ profiles remain outside the scope of the present photospheric fit.

\begin{acknowledgements}
M.C., I.A. and C.A. thank FONDECYT projects 1230131 and 1261498 for supporting this study. M.C. and C.A. acknowledge partial support from the Centro de Astrofísica de Valparaíso. D.T. acknowledges support from ANID BECAS/DOCTORADO NACIONAL/2024-21240465. G.-W. Li acknowledges the support from the National Key R\&D Program of China No.2025YFF0511004. This work was partially co-funded by the European Union (Project 101183150-OCEANS). This study was made possible by the use of AWS-U. Chile-NLHPC credits. Powered@NLHPC: This research was partially supported by the NLHPC's supercomputing infrastructure (CCSS210001).
Finally, we thank the referee, J. Zorec, for his valuable comments in improving this study.
\end{acknowledgements}
\bibliographystyle{aa}
\bibliography{referencesV5}

\begin{thebibliography}{24}
\expandafter\ifx\csname natexlab\endcsname\relax\def\natexlab#1{#1}\fi

\bibitem[{{de Mink} {et~al.}(2013){de Mink}, {Langer}, {Izzard}, {Sana}, \& {de
  Koter}}]{deMink2013}
{de Mink}, S.~E., {Langer}, N., {Izzard}, R.~G., {Sana}, H., \& {de Koter}, A.
  2013, \apj, 764, 166

\bibitem[{{Domiciano de Souza} {et~al.}(2014){Domiciano de Souza}, {Kervella},
  {Moser Faes}, {Dalla Vedova}, {M\'erand}, {Le Bouquin}, {Espinosa Lara},
  {Rieutord}, {Bendjoya}, {Carciofi}, {Hadjara}, {Millour}, \&
  {Vakili}}]{Domiciano2014}
{Domiciano de Souza}, A., {Kervella}, P., {Moser Faes}, D., {et~al.} 2014,
  A\&A, 569, A10

\bibitem[{{Domiciano de Souza} {et~al.}(2002){Domiciano de Souza}, {Vakili},
  {Jankov}, {Janot-Pacheco}, \& {Abe}}]{2002A&A...393..345D}
{Domiciano de Souza}, A., {Vakili}, F., {Jankov}, S., {Janot-Pacheco}, E., \&
  {Abe}, L. 2002, \aap, 393, 345

\bibitem[{{Espinosa-Lara} \& {Rieutord}(2011)}]{EspinosaLaraRieutord2011}
{Espinosa-Lara}, F. \& {Rieutord}, M. 2011, \aap, 533, A43

\bibitem[{{Fr{\'e}mat} {et~al.}(2005){Fr{\'e}mat}, {Zorec}, {Hubert}, \&
  {Floquet}}]{Fremat2005}
{Fr{\'e}mat}, Y., {Zorec}, J., {Hubert}, A.-M., \& {Floquet}, M. 2005, \aap,
  440, 305

\bibitem[{{Hadjara} {et~al.}(2014){Hadjara}, {Domiciano de Souza}, {Vakili},
  {Jankov}, {Millour}, {Meilland}, {Khorrami}, {Chelli}, {Baffa}, {Hofmann},
  {Lagarde}, \& {Robbe-Dubois}}]{2014A&A...569A..45H}
{Hadjara}, M., {Domiciano de Souza}, A., {Vakili}, F., {et~al.} 2014, \aap,
  569, A45

\bibitem[{{Hadjara} {et~al.}(2022){Hadjara}, {Petrov}, {Jankov},
  {Cruzal{\`e}bes}, {Boskri}, {Spang}, {Lagarde}, {He}, {Chen}, {Nitschelm},
  {de Almeida}, {Pereira}, {Michael}, {Gao}, {Wang}, {Reyes}, {Arcos}, {Araya},
  \& {Cur{\'e}}}]{2022MNRAS.511.4724H}
{Hadjara}, M., {Petrov}, R.~G., {Jankov}, S., {et~al.} 2022, \mnras, 511, 4724

\bibitem[{{Hadjara} {et~al.}(2018){Hadjara}, {Petrov}, {Jankov},
  {Cruzal{\`e}bes}, {Spang}, \& {Lagarde}}]{2018MNRAS.480.1263H}
{Hadjara}, M., {Petrov}, R.~G., {Jankov}, S., {et~al.} 2018, \mnras, 480, 1263

\bibitem[{{Hubeny} \& {Lanz}(1995)}]{HubenyLanz1995}
{Hubeny}, I. \& {Lanz}, T. 1995, \apj, 439, 875

\bibitem[{{Hubeny} \& {Lanz}(2000)}]{HubenyLanz2000}
{Hubeny}, I. \& {Lanz}, T. 2000, {SYNSPEC---A User's Guide, Version 43,
  NASA/GSFC}

\bibitem[{{Hubeny} \& {Lanz}(2011)}]{HubenyLanz2011}
{Hubeny}, I. \& {Lanz}, T. 2011, {SYNSPEC: General Spectrum Synthesis Program},
  Astrophysics Source Code Library, record ascl:1109.022

\bibitem[{{Hubeny} \& {Lanz}(2017)}]{HubenyLanz2017}
{Hubeny}, I. \& {Lanz}, T. 2017, arXiv e-prints, arXiv:1706.01859

\bibitem[{{Levenhagen}(2014)}]{Levenhagen2014}
{Levenhagen}, R.~S. 2014, \apj, 797, 29

\bibitem[{{Levenhagen} {et~al.}(2024){Levenhagen}, {Cur{\'e}}, {Arcos}, {Diaz},
  {Araya}, {Am{\^o}res}, {Turis-Gallo}, \& {Concha}}]{Levenhagen2024}
{Levenhagen}, R.~S., {Cur{\'e}}, M., {Arcos}, C., {et~al.} 2024, \aap, 685, A57

\bibitem[{{Li}(2020)}]{Li2020}
{Li}, G.-W. 2020, \apjl, 892, L26

\bibitem[{{Packet}(1981)}]{Packet1981}
{Packet}, W. 1981, \aap, 102, 17

\bibitem[{{Rieutord}(2016)}]{Rieutord2016}
{Rieutord}, M. 2016, in Lecture Notes in Physics, Vol. 914, Cartography of the
  Sun and the Stars (Springer), 101

\bibitem[{{Sana} {et~al.}(2012){Sana}, {de Mink}, {de Koter}, {Langer},
  {Evans}, {Gieles}, {Gosset}, {Izzard}, {Le Bouquin}, \&
  {Schneider}}]{Sana2012}
{Sana}, H., {de Mink}, S.~E., {de Koter}, A., {et~al.} 2012, Science, 337, 444

\bibitem[{{Sch{\"o}ning} \& {Butler}(1989)}]{SchoeningButler1989}
{Sch{\"o}ning}, T. \& {Butler}, K. 1989, \aaps, 78, 51, erratum: 1989, A\&AS,
  79, 153

\bibitem[{{Stoeckley}(1968)}]{Stoeckley1968}
{Stoeckley}, T.~R. 1968, \mnras, 140, 141

\bibitem[{{Townsend} {et~al.}(2004){Townsend}, {Owocki}, \&
  {Howarth}}]{Townsend2004}
{Townsend}, R.~H.~D., {Owocki}, S.~P., \& {Howarth}, I.~D. 2004, \mnras, 350,
  189

\bibitem[{{Turis-Gallo} {et~al.}(2025){Turis-Gallo}, {Cur{\'e}}, {Levenhagen},
  {Arcos}, {Araya}, \& {Christen}}]{Turis-Gallo2025}
{Turis-Gallo}, D., {Cur{\'e}}, M., {Levenhagen}, R.~S., {et~al.} 2025, \aap,
  695, A129

\bibitem[{{Vidal} {et~al.}(1970){Vidal}, {Cooper}, \&
  {Smith}}]{VidalCooperSmith1970}
{Vidal}, C.~R., {Cooper}, J., \& {Smith}, E.~W. 1970, Journal of Quantitative
  Spectroscopy and Radiative Transfer, 10, 1011

\bibitem[{{von Zeipel}(1924)}]{vonZeipel1924}
{von Zeipel}, H. 1924, \mnras, 84, 665

\end{thebibliography}

\end{document}